# Generator of Neural Network Potential for Molecular Dynamics: Constructing Robust and Accurate Potentials with Active Learning for Nanosecond-Scale Simulations

*Naoki Matsumura[1]\*, Yuta Yoshimoto[1], Tamio Yamazaki[2], Tomohito Amano[3], Tomoyuki Noda[4], Naoki Ebata[5], Takatoshi Kasano[4] and Yasufumi Sakai[1]*

[1] Fujitsu Research, Fujitsu Limited, 4-1-1, Kamiodanaka, Nakahara-ku, Kawasaki, Kanagawa 211-8588, Japan

[2] JSR-UTokyo Collaboration Hub, CURIE, JSR Corporation, 1-9-2, Higashi-Shinbashi, Minato-ku, Tokyo 105-8640, Japan

[3] Department of Physics, The University of Tokyo, Hongo, Bunkyo-ku, Tokyo 113-0033, Japan

[4] Advanced Technology Services Business Unit, Fujitsu Limited, 1-5, Omiyacho, Saiwai-ku, Kawasaki, Kanagawa 212-0014, Japan

[5] Public Business Unit, Fujitsu Limited, 1-5, Omiyacho, Saiwai-ku, Kawasaki, Kanagawa 212-0014, Japan

\*Email: matsumura-naoki@fujitsu.com

**Abstract**

Neural network potentials (NNPs) enable large-scale molecular dynamics (MD) simulations of systems containing >10,000 atoms with the accuracy comparable to *ab initio* methods and play a crucial role in material studies. Although NNPs are valuable for short-duration MD simulations, maintaining the stability of long-duration MD simulations remains challenging due to the uncharted regions of the potential energy surface (PES). Currently, there is no effective methodology to address this issue. To overcome this challenge, we developed an automatic generator of robust and accurate NNPs based on an active learning (AL) framework. This generator provides a fully integrated solution encompassing initial data set creation, NNP training, evaluation, sampling of additional structures, screening, and labeling. Crucially, our approach uses a sampling strategy that focuses on generating unstable structures with short interatomic distances, combined with a screening strategy that efficiently samples these configurations based on interatomic distances and structural features. This approach greatly enhances the MD simulation stability, enabling nanosecond-scale simulations. We evaluated the performance of our NNP generator in terms of its MD simulation stability and physical properties by applying it to liquid propylene glycol (PG) and polyethylene glycol (PEG). The generated NNPs enable stable MD simulations of systems with >10,000 atoms for 20 ns. The predicted physical properties, such as the density and self-diffusion coefficient, show excellent agreement with the experimental values. This work represents a remarkable advance in the generation of robust and accurate NNPs for organic materials, paving the way for long-duration MD simulations of complex systems.

# 1. Introduction

In recent years, neural network potentials (NNPs) have shown promise as a tool for dramatically enhancing the computational efficiency of *ab initio* or density functional theory (DFT)[1] calculations while preserving accuracy comparable to them. NNPs are trained to learn the complex mapping between an atomic structure and its potential energy surface (PES). By accurately reproducing the PES, NNPs have opened new avenues for research and application, providing unprecedented insights into complex materials and molecular systems[2–4]. This advancement has enabled researchers to investigate phenomena that were previously computationally prohibitive, including long-time and large-scale` molecular dynamics (MD) simulations with DFT accuracy. To further elevate the predictive accuracy of the PES, a wide variety of NNP models, such as DeepPot-SE[5], PaiNN[6], QRNN[7], M3GNet[8], and CHGNet[9], have been developed.

Although NNP architectures have become more sophisticated, their performance remains heavily dependent on the quality and size of the training data sets. Generally, larger training data sets lead to better performance[10–13]; however, the data labeling process, which involves calculating the energy and atomic forces, is computationally very expensive. To reduce the quantity of data to be labeled, active learning (AL) techniques are being actively researched[14–23]. AL aims to iteratively augment the training data set with diverse structures, focusing on regions of the configuration space where the NNP model exhibits poor predictive accuracy. The AL process begins by training an initial NNP model using a small initial data set generated from methods such as *ab initio* molecular dynamics (AIMD) simulations and normal mode sampling[10]. Additional structures were then collected using sampling methods such as MD simulations with the NNP model (NNP-MD). To avoid redundant labeling calculations, the additional structures were screened to prioritize those dissimilar to the existing data set and capable of improving the current

NNP model based on uncertainty criteria. These screened structures were subsequently labeled using *ab initio* calculations. The NNP model was then retrained to improve its fitting accuracy with the enhanced data set, and its performance was evaluated. This data generation and refinement process with AL can be repeated until the accuracy of the NNP is deemed sufficient. As these processes indicate, developing high-accuracy NNPs requires a substantial effort. We note that the pretrained advanced NNPs, such as MACE-MP-0[24] and MatterSim[25], may require only a small amount of training data to achieve desirable accuracy and robustness. This is because they are already possess sufficient accuracy and robustness to perform NNP-MD simulations[25–27]. However, these studies also suggest that AL is still necessary for more complex systems or when targeting different levels of DFT accuracy.

Although promising results using NNPs constructed by AL have been reported for short-duration simulations[17,18,22,23], it remains unclear whether these potentials can reliably predict long-time dynamics and maintain stability over extended nanosecond-scale simulation periods. This is a crucial consideration for practical applications, particularly for materials such as glass and polymers. Behler[28] and Stocker et al.[11] have discussed the stability in terms of the presence of holes in the PES and the possibility of the NNP-MD simulations falling into these holes. They further concluded that a thorough discussion of stability requires the execution of long-time NNP-MD simulations, as short-term simulations may not reveal potential issues. Fu et al.[29] conducted a study investigating the relationships among model architectures and the stability of MD simulations for organic and inorganic materials. Their findings revealed no strong correlation between the accuracy of the force predictions and the stability of the MD simulations. This suggests that improving the stability of MD simulations may require focusing on the design of the data sets rather than the model architecture optimization. Although these studies discuss the issue of MD simulation stability and its underlying causes, they lack a unified methodology or solution for addressing these challenges. Chahal et al.[30] have developed an accurate NNP capable of performing nanoscale NNP-MD simulations for polyacrylonitrile polymers. However, the data set used in their study was manually selected from the trajectories of the AIMD simulations, resulting in high computational labeling costs. Furthermore, this manual selection process introduces a major bottleneck, as it is time-consuming and prone to human error, thereby affecting the overall efficiency and reliability of the NNP development. Consequently, there is a pressing need for more automated and systematic approaches to generating robust and accurate NNPs.

In this work, we developed an automatic NNP generator employing an efficient AL scheme, designed to generate robust and accurate NNPs for any chemical system. It explores configuration spaces using NNP-MD simulations and selects structures leading to the improvement of the current NNP while ensuring data set diversity through a two-step screening phase using model uncertainty-based and structural feature-based methods. In addition, it addresses the stability issue of NNP-MD with an efficient sampling method to obtain unstable structures, together with an enhanced structural feature-based screening method that considers structural features and distance metrics. We will initially explain the workflow. Next, we will show the results of applying this scheme to liquid propylene glycol (PG) and polyethylene glycol (PEG), along with discussions on the simulation stability. This work paves the way for the development of robust and highly accurate NNPs for a wide range of chemical systems, with particular emphasis on organic materials.

## 2. Generator of Neural Network Potential

In this section, we introduce our NNP generator, which aims to generate a robust and accurate NNP via AL by employing efficient sampling and screening methods for the initial structure inputs. The data set is composed of structures, along with their corresponding energies, forces, and stresses calculated using DFT. Its AL workflow, initiated after an initial data set creation phase, is composed of the following consecutive phases: NNP training, NNP evaluation, sampling, screening, and labeling, as illustrated in **Figure 1**. These phases are conducted iteratively, continuously refining the NNP model until a predetermined number of iterations is reached. Each phase is explained in detail below.

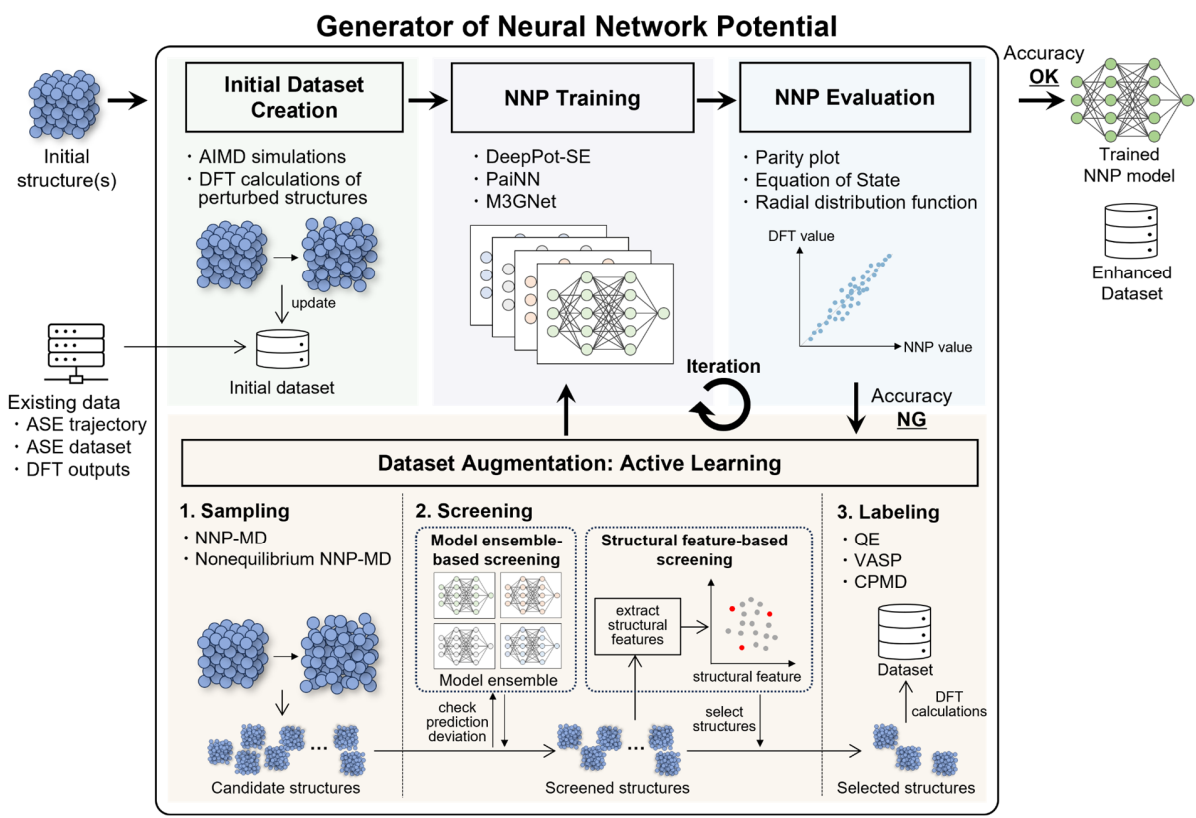

**Figure 1.** Overview of our NNP generator. The process begins with the initial data set creation for the initial structures. This is followed by the AL cycle, which includes NNP training, NNP evaluation, sampling, screening, and labeling. This cycle repeats to iteratively refine the model.

*Initial Data Set Creation*: AL processes are initiated with the creation of an initial data set to train an initial NNP model. For the initial structures, our NNP generator creates the initial data set using either or both of the following methods: AIMD simulations and random displacement sampling, the latter of which generates structures by randomly perturbing the initial configurations. These methods were performed using the Atomic Simulation Environment (ASE) package[31]. The DFT packages available within this generator are VASP[32,33], Quantum ESPRESSO[34–36], and CPMD[37]. Additionally, the generator can load existing data, such as DFT calculation results or user-prepared data sets, thereby enabling the use of any data set for AL. The supported formats include output files from the DFT packages, ASE trajectory files, and ASE data set files.

*NNP Training*: This phase involves training the NNP models on the data set. Several models with an identical architecture but different initialization seeds are trained to estimate the uncertainty criteria (description of the screening phase). The model initialized with the first seed (referred to as model 0) is employed in the subsequent evaluation and sampling phases, while all models are used in the screening phase. Our NNP generator currently supports DeepPot-SE[5] (implemented in the DeePMD-kit[38,39]), PaiNN[6], and M3GNet[8]. Users can choose to train a model from scratch or use a model pretrained on data sets such as the Open Catalyst Project data sets[12,13] for PaiNN, the Materials Project data set[40] for M3GNet, or any user-defined data set. The training can be restarted from a previous iteration's model with a specified small number of training steps to be more effective. In addition to these options, our NNP generator provides flexibility in hyperparameter tuning, enabling users to optimize the training process for their specific needs. The generator also supports parallel and distributed computing, minimizing the time required for NNP training.

***NNP Evaluation*:** This phase evaluates the trained NNP model to verify its accuracy and reliability. One of the primary evaluation methods is the comparison of the NNP model's output results with those obtained from DFT. This comparison measures the accuracy of the energy and force predictions using metrics such as the mean absolute error and the root mean squared error (RMSE). Another evaluation method is the analysis of the equation of state, which examines the relationship between energy and volume. Additionally, the radial distribution function analysis was employed to evaluate the system structural properties.

***Sampling*:** In this phase, NNP-MD simulations using the trained NNP model were conducted to explore the configuration space and generate a set of candidate structures. Our NNP generator supports NNP-MD simulations in the constant temperature, constant volume (NVT) and constant pressure, constant temperature (NPT) ensembles, allowing for the exploration of a configuration space by varying the temperatures and pressures. Additionally, it supports nonequilibrium MD simulations using the NNP model (NNP-NEMD) to capture unstable structures[41]. The NNP-NEMD continually changes the simulation box volume and the atomic positions relative to the volume during the dynamics run, providing highly compressed or expanded systems with short or long atomic distances. Our NNP generator uses the Large-scale Atomic/Molecular Massively Parallel Simulator (LAMMPS) package[42] for DeepPot-SE and the ASE package[31] for PaiNN and M3GNet. Currently, the NNP-NEMD simulations are only supported in LAMMPS. By leveraging these simulations, our NNP generator can effectively sample a wide range of configurations, ensuring that the final NNP model is well-trained and capable of accurate predictions under the specified target conditions.

***Screening*:** This phase reduces the number of candidate structures passed to the labeling phase. The primary objective is to identify the candidate structures that are not present in the existing data set and contribute to the improvement of the current NNP model. To achieve this, we adopted a combination of two screening methods: model ensemble-based[14,17] and structural feature-based[18,43,44] methods. Initially, we apply the model ensemble-based screening method to filter candidates based on model prediction accuracy. Next, we use the structural feature-based screening method to refine the selection based on structural features. The former method reduces the number of candidates, while the latter narrows it down further to a user-defined manageable number.

The model ensemble-based screening method uses the query-by-committee[45] (QBC) approach. QBC measures the disagreement among multiple models within a committee (ensemble) for a given candidate structure, serving as an indicator of structures that are not well-trained. Here, we employed four models with the same NNP architecture but with different initial weights for the model ensembles. We adopted the maximum model deviation indicator[17,39], defined as the maximum standard deviation of the atomic forces predicted by the model ensemble, as follows:

$$\epsilon(x) = \max\left\{\sqrt{\langle \|F_i(x; \theta_k) - \langle F_i(x; \theta_k)\rangle\|^2\rangle} \,|\, 1 \le i \le N\right\} \qquad (1)$$

where $x$, $F_i$, $\theta_k$, and $N$ denote a certain data frame containing the coordinates and chemical species of all atoms, the atomic force on atom $i$, the parameters of the model $k$, and the number of atoms in the system, respectively. The ensemble average, $\langle \cdot \rangle$, is calculated over $n_\mathrm{m}$ models as:

$$\langle F_i(x; \theta_k)\rangle = \frac{1}{n_\mathrm{m}} \sum_{k=1}^{n_\mathrm{m}} F_i(x; \theta_k). \qquad (2)$$

A lower value of $\epsilon(x)$ indicates that the given candidate structure is well-represented; therefore, there is no need to add it to the data set. Conversely, a larger value of $\epsilon(x)$ suggests that the candidate structure is poorly represented or considered a failed structure. Thus, the preferred structures to be labeled are within the middle of the range[17]. This model ensemble-based screening method calculates $\epsilon(\boldsymbol{x})$ for all candidate structures and selects those within the user-defined range. Based on our experience, a range of 0.05 to 0.20 eV/Å generally works well across typical organic systems.

The structural feature-based screening method extracts intermediate features from the NNP model and projects them onto a two-dimensional (2D) space using a dimensionality reduction method. This projection identifies the candidate structures that are most distinct from the structures present in the current data set. The intermediate features are derived from the outputs of the NNP's descriptor layers, which dynamically change through AL iterations due to the presence of trainable parameters. Consequently, the intermediate features derived from the same structure will be updated in each iteration. This dynamic procedure identifies candidate structures that are most beneficial to the current NNP. Previous works[18,43,44] used predefined symmetry functions[46] to encode structural information as features. The nonadaptive nature of the predefined symmetry functions may limit their screening ability for the current NNP model enhancements. Our NNP generator implements several dimensionality reduction methods, including PCA[47], t-SNE[48], UMAP[49] and densMAP[50]. After visualizing the structural features in a 2D space, all Euclidean distances between the candidate structures were calculated. For each pair of structures with short Euclidean distances, one structure is removed in order of distance until a predefined number of candidates remain. Additionally, our NNP generator implements a screening function that can select based on structural features and specific quantities such as energy, maximum force, volume, and minimum interatomic distance. This is achieved by extending the 2D space into a three-dimensional (3D) space through the incorporation of the specific quantity. By leveraging these screening techniques, our NNP generator ensures that the most informative and diverse candidate structures are selected for the labeling phase, thereby enhancing the robustness and accuracy of the NNP model.

*Labeling*: In this phase, the reference energies, forces, and stresses for the selected candidate structures are calculated using DFT. Our NNP generator supports several DFT packages, including VASP[32,33], Quantum ESPRESSO[34–36], and CPMD[37]. After the DFT calculations, the labeled structures were then filtered based on a user-defined threshold for the absolute magnitude of the maximum force. The remaining labeled structures were split into training and validation data sets according to a predefined ratio. The DFT settings can be configured for each structure individually, allowing the use of different DFT settings within the same AL run. This flexibility manages the diverse computational requirements and optimizations.

## 3. Results and Discussion

In this section, we investigate the capability of the models generated by our NNP generator for liquid PG and PEG. Initially, we describe the generator settings. Next, we present the generator execution results and the capabilities of the generated models for each system. **Figure 2** shows the various configurations of PG and PEG used in this study.

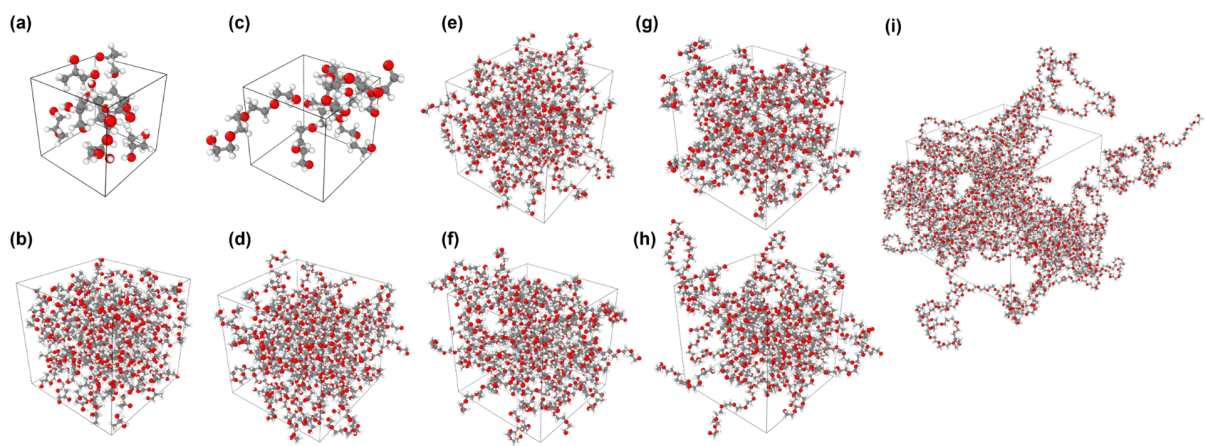

**Figure 2.** PG systems containing **(a)** 10 molecules (130 atoms) for data creation and **(b)** 216 molecules (2808 atoms) for production runs. PEG systems containing **(c)** five 4-mer molecules (155 atoms) for data creation, and **(d)** a hundred 4-mer molecules (3100 atoms), **(e)** eighty 5-mer molecules (3040 atoms), **(f)** seventy 6-mer molecules (3150 atoms), **(g)** sixty 7-mer molecules (3120 atoms), **(h)** fifty 8-mer molecules (2950 atoms), and **(i)** ten 150-mer molecules (10,530 atoms) for production runs. These structures were visualized using the OVITO package[51].

### 3.1. Propylene Glycol (PG)

#### 3.1.1. Settings of the NNP Generator

We used DeePMD-kit 2.2.4[38,39] for training, LAMMPS 2Aug2023[42] for sampling, densMAP[50] (implemented in umap_learn 0.5.6[49]) for screening, and CPMD 4.2[37] for labeling. To initiate the process, we prepared an initial structure consisting of 10 PG molecules (130 atoms total) randomly placed using the Packmol package[52], as shown in **Figure 2(a)**. The density was fixed at 1.04 g/cm$^3$, corresponding to the experimental value[53]. An equilibration simulation of 1 ns was then performed in the NVT ensemble at 300 K, employing velocity rescaling with a 1 fs time step, using the GROMACS package[54]. The general AMBER force field version 2 (GAFF2)[55] and the AM1-BCC charge[56] were applied. Topology files were generated by the Antechamber package[57] and the ACPYPE package[58]. Periodic boundary conditions were applied in all the simulations.

*Initial Data Set Creation*: For the initial data set generation, AIMD simulations were conducted using the NVT ensemble with the Nosé–Hoover thermostat[59,60]. Simulations were performed at temperatures of 300 and 600 K, each for 1 ps with a time step of 1 fs, generating a total of 2000 structures. The data set was split into training and validation sets in an 8:2 ratio.

*NNP Training*: The se_e2_a descriptor of DeepPot-SE[5] was employed with a cutoff radius of 6 Å. The embedding and fitting networks were composed of (25, 50, 100) and (120, 120, 120) nodes with ResNet-like architectures[61], respectively. Four models, each initialized with different random seeds, were trained to calculate the maximum model deviation of atomic forces, $\epsilon(\boldsymbol{x})$. To enhance the training efficiency, the trained models were inherited from the previous iteration. These models were trained for 400,000 steps on the initial data set (iteration 0), while they were trained for 200,000 at subsequent iterations. The Adam optimizer[62] was used. The learning rate decayed from $1.0 \times 10^{-3}$ to $1.0 \times 10^{-8}$ with 5000 decay steps. After generating all the training data, a final model was trained for 500,000 steps using a data parallel training approach with four processes (equivalent to 2,000,000 steps in serial training). The fitting network size of the final model was increased to (240, 240, 240).

*Sampling*: For iteration 1 to 10, NNP-MD simulations of 100 ps with a time step of 1 fs were performed using the NVT and NPT ensembles, with temperatures ranging from 300 to 600 K in 50 K increments and a pressure of 1 bar, thereby generating 14 trajectories in each iteration. The Nosé–Hoover thermostat[59,60] was used for both ensembles, and the Parrinello–Rahman barostat[63] was applied in the NPT ensemble. Candidate structures were extracted every 100 fs. For iteration 11, we performed NNP-NEMD simulations to generate unstable structures. This sampling strategy together with the 3D structural feature-based screening method significantly improves the stability of the NNP-MD simulations, as detailed in Section 3.1.3.

*Screening*: For iteration 1 to 10, the acceptable range of $\epsilon(\boldsymbol{x})$ was set to 0.05 to 0.15 eV/Å, determined from four trained models with the same model architecture but with different initial seeds. The densMAP method[50] was used for dimensionality reduction. The maximum number of structures for the DFT calculations was set to 1000 per iteration. For iteration 11, we used the 3D structural feature-based screening to select the structures which have short O–H distances, as detailed in Section 3.1.3.

*Labeling*: All DFT calculations were performed using the CPMD package[37]. The Becke–Lee–Yang–Parr (BLYP) exchange-correlation functional[64,65] was used. The Goedecker–Teter–Hutter (GTH) pseudopotentials[66] were employed with a plane-wave cutoff of 100 Ry. To include the dispersion, an empirical D2 dispersion correction[67] was applied.

**3.1.2. Results of the NNP Generator Execution**

We initially present the execution results of the AL. **Table 1** details the iterative construction of the training data set. Over 10 iterations, a total of 9997 labeled structures were generated. As described below, iteration 11 was conducted to improve the stability of the NNP-MD simulations, resulting in the generation of 500 labeled structures. The number of structures screened by the model ensemble-based screening method gradually increased as the NNP's predictive capability improved, enabling it to accurately infer a wide range of structures. Even as the number of structures screened by the model ensemble-based screening method increases, the structural feature-based screening method can reduce the number of structures to a predetermined size base on the similarities in the structural features. Notably, out of 10,500 DFT calculations performed, only three failed, demonstrating the effectiveness of the model ensemble-based screening method in filtering out unrealistic structures using an upper threshold of $\epsilon(\boldsymbol{x})$.

**Table 1.** Construction of the training data set for PG[a]

| iter | sampling | | | | screening | | labeling |
|---|---|---|---|---|---|---|---|
| | method | ensemble | temperature (K) | # of sampled structures | # of structures selected by the model ensemble-based method | # of structures selected by the structural feature-based method | # of labeled structures |
| 0 | AIMD | NVT | 300, 600 | 2000 | | | 2000 |
| 1 | NNP-MD | NVT, NPT | 300–600 | 10,658 | 1323 | 1000 | 1000 |
| 2 | NNP-MD | NVT, NPT | 300–600 | 13,335 | 3240 | 1000 | 998 |
| 3 | NNP-MD | NVT, NPT | 300–600 | 13,557 | 8992 | 1000 | 1000 |
| 4 | NNP-MD | NVT, NPT | 300–600 | 14,000 | 9371 | 1000 | 999 |
| 5 | NNP-MD | NVT, NPT | 300–600 | 14,000 | 8730 | 1000 | 1000 |
| 6 | NNP-MD | NVT, NPT | 300–600 | 13,386 | 9675 | 1000 | 1000 |
| 7 | NNP-MD | NVT, NPT | 300–600 | 14,000 | 10,100 | 1000 | 1000 |
| 8 | NNP-MD | NVT, NPT | 300–600 | 13,433 | 10,952 | 1000 | 1000 |
| 9 | NNP-MD | NVT, NPT | 300–600 | 13,608 | 11,403 | 1000 | 1000 |
| 10 | NNP-MD | NVT, NPT | 300–600 | 13,256 | 10,544 | 1000 | 1000 |
| 11 | NNP-NEMD (compress to 70%) | | 300–700 | 10,000 | 7066 | 500 | 500 |

[a]Each row details the additions to the training data set at each iteration. The columns represent iteration, sampling method, ensemble, temperature, number of sampled structures, number of screened structures by the model ensemble-based method, number of screened structures by the structural feature-based method, and number of labeled structures. The pressure of NNP-MD simulations in the NPT ensemble was set to 1 bar.

**Figure 3(a)** and **(b)** show the 2D feature spaces of the structures in the data sets before and after the structural feature-based screening at iteration 10, respectively. The results for the other iterations are presented in **Figure S1** in the Supporting Information. Our NNP generator effectively selected structures that did not overlap with the existing labeled structures, demonstrating its capability to collect diverse and representative data. This strategic selection process enhances the generalizability of the NNP model by avoiding redundancy and focusing on the underrepresented areas of the feature space.

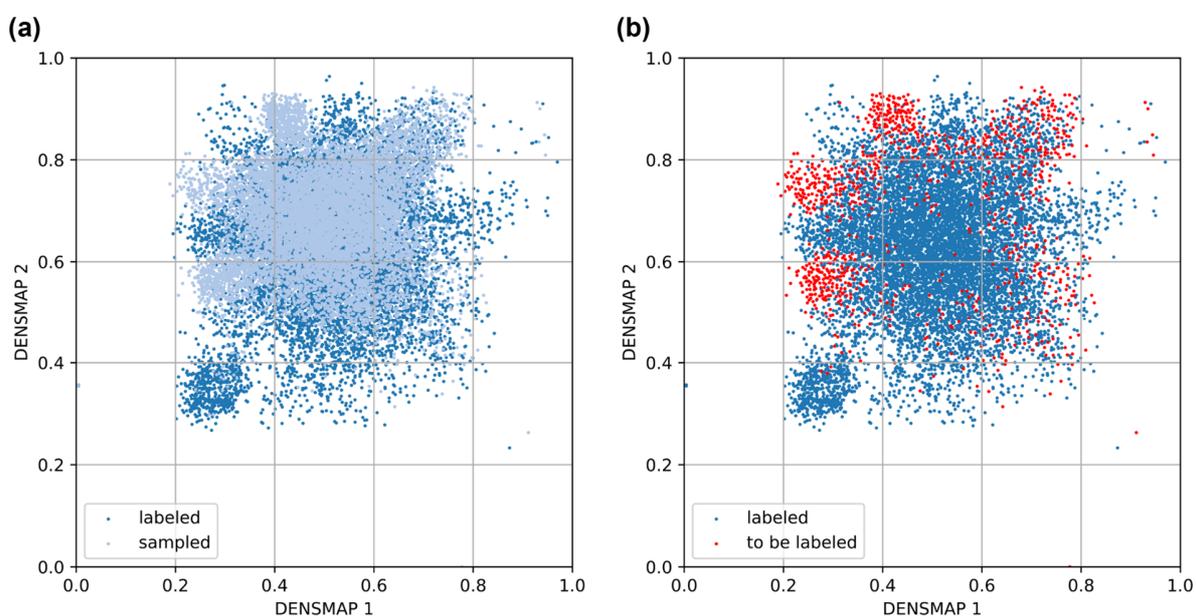

**Figure 3.** 2D feature space representation of the structures **(a)** before and **(b)** after applying the structural feature-based screening method at iteration 10. The dark blue, light blue, and red points represent structures labeled up to iteration 9, structures sampled at iteration 10, and structures selected for labeling, respectively.

To assess the predictive accuracy of the model, we calculated the RMSE for the energy and force between DFT and NNP (**Figure S2**). The validation RMSEs of the first model trained on the AIMD training data set were 1.08 meV/atom for energy and 86.5 meV/Å for force. After training on the data set at iteration 10, these values changed to 2.71 meV/atom and 61.6 meV/Å, respectively. The slight increase in the energy error can be attributed to the broader energy distribution that AL sampled. This distribution is illustrated in **Figure 4(a)**. The initial energy histogram shows two distinct peaks corresponding to AIMD results at 300 and 600 K. As iterations progressed, additional data points were generated to fill the gap between these peaks. Meanwhile, the force distribution, as illustrated in **Figure 4(b)**, is centered around 0 eV/Å and extends to ±4 eV/Å.

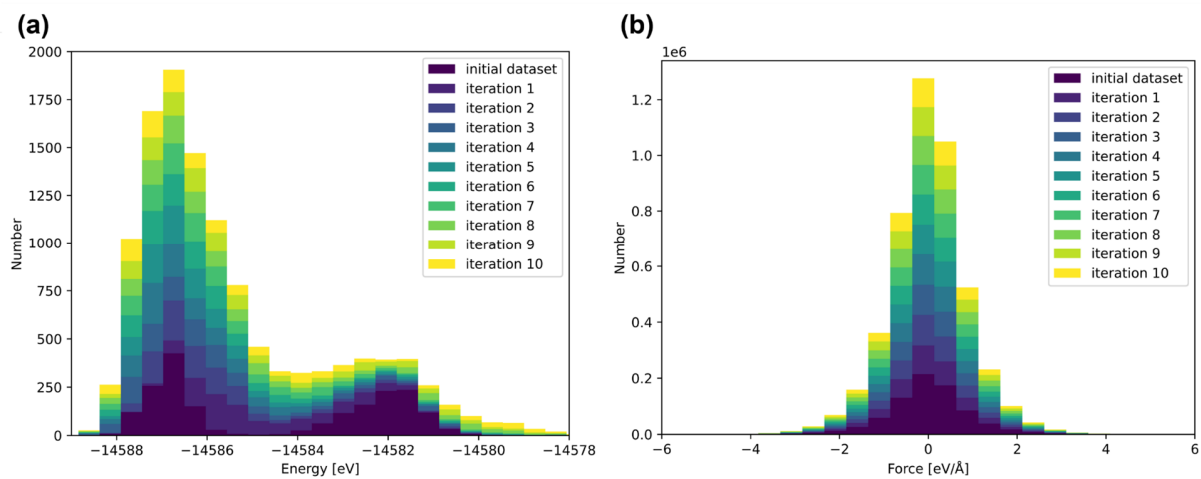

**Figure 4**. Histograms of **(a)** energy and **(b)** force in the data set over 10 iterations. The energy histogram of the initial data set (dark blue bars) exhibits two distinct peaks corresponding to AIMD results at 300 K (from −14,588 to −14,586 eV) and 600 K (from −14,584 to −14,580 eV).

### 3.1.3. Running the Long-Time NNP-MD Simulations

To evaluate the accuracy of the generated model, we performed long-time simulations on a large system comprising 216 PG molecules (2808 atoms), greatly exceeding the number of atoms in the training data set (130 atoms). The initial configuration was created using Packmol[52,68] to roughly pack them. The molecules were subsequently compressed to a density of 1.04 g/cm$^3$ using an NNP-MD simulation with the generated model in the NPT ensemble at 293.15 K and 10,000 bar. This compressed structure was the starting point for the subsequent NNP-MD simulations.

The stability of the NNP-MD simulations was evaluated using the maximum model deviation indicator[17,39]. Four independent NNP models, trained with different random seeds (referred to as model 0 through model 3), were employed. Each model was used to run the NNP-MD simulation at 293.15 K and 1 bar for 1 ns for equilibration and for 20 ns for the production run, with time steps of 0.5 fs, generating four distinct trajectories. The maximum model deviation was then calculated for each trajectory using all four models. A substantial increase indicates that the simulated structures are underrepresented in the training data set, suggesting that the NNP-MD simulations are unstable. The key properties, such as density, self-diffusion coefficient, thermal expansion coefficient, and isothermal compressibility, were calculated from the production run trajectories.

**Figure 5(a)** presents the densities and maximum model deviations of the NNP-MD simulations using the four models trained on data up to iteration 10. During the NNP-MD simulation of model 1, the system collapsed at approximately 18 ps due to the unphysical separation of a hydrogen atom from the molecule. **Figure 6** illustrates the process of this collapse. At 18.2 ps, the intermolecular distance decreased. Then, from 18.3 ps to 18.4 ps, the two hydroxy groups attracted each other, and one of the hydrogen atoms oscillated between the oxygen atoms. Finally, the hydrogen atom of the molecule in the lower left corner detached from its parent molecule, triggering the simulation collapse. This phenomenon is caused by unrealistic forces (specifically, 6.0 eV/Å for the H atom) generated when the O–H distances become too short.

Notably, the O–H distance between the molecules just before the hydrogen atom detached was approximately 0.81 Å. This indicates that the NNP model may require further refinement or additional training data to improve the representation of the PES, as these unrealistic forces are expected to stem from an inadequate representation of the PES.

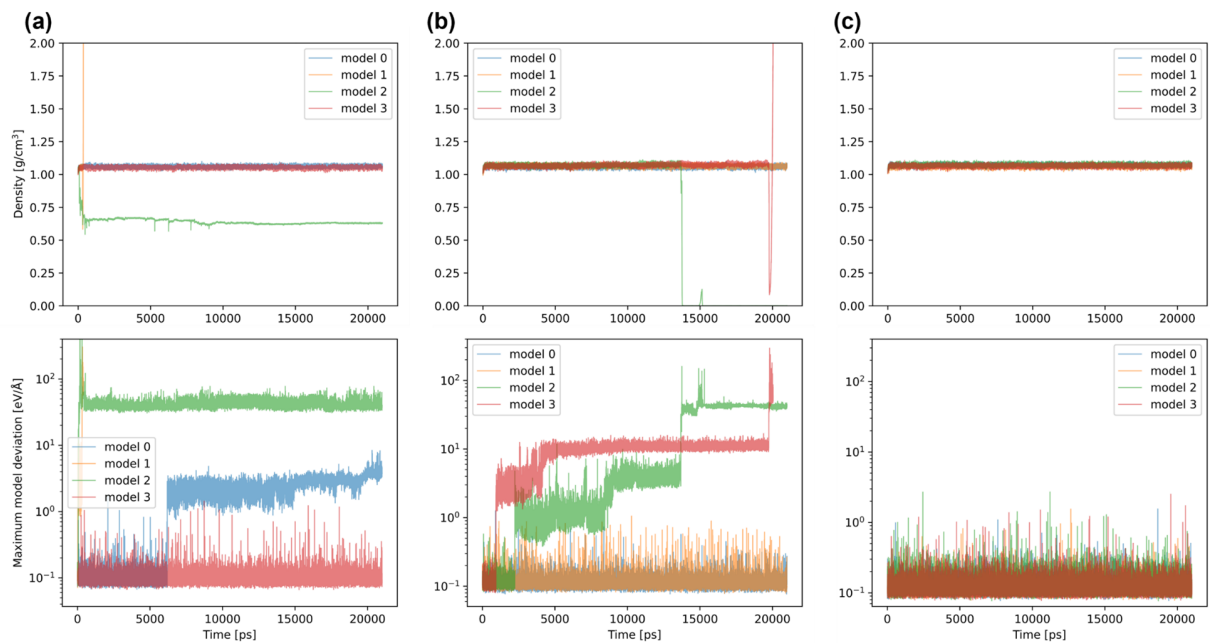

**Figure 5.** Densities (top) and maximum model deviations of atomic forces (bottom) of the simulations for PG at 293.15 K and 1 bar using four NNPs generated at **(a)** iteration 10, **(b)** iteration 11 with the 2D structural feature-based screening, and **(c)** iteration 11 with the 3D structural feature-based screening.

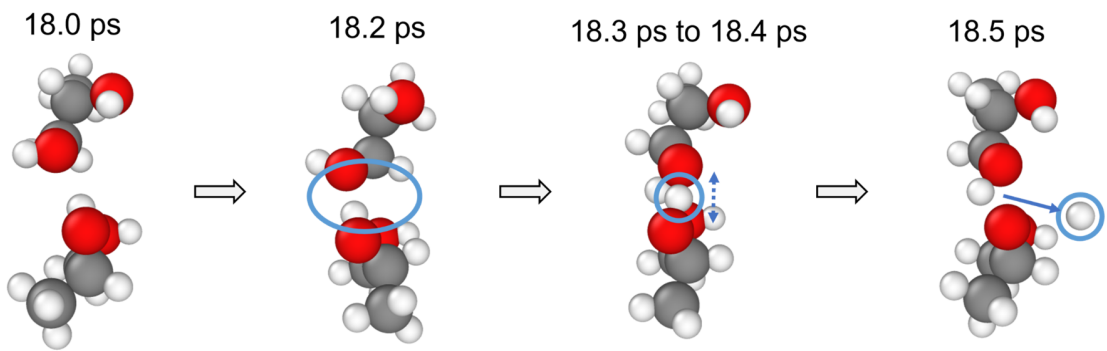

**Figure 6.** Collapse process in the liquid PG system. To visualize this phenomenon, only two molecules from the entire system are shown. At 18.2 ps, the two molecules come close together. From 18.3 to 18.4 ps, a hydrogen atom in one molecule is attracted to the other molecule and oscillates between the two. Finally, at 18.5 ps, the hydrogen atom separated from the molecule. This process was visualized using the OVITO package[51].

To further improve the current NNP model, we conducted iteration 11 to collect structures with short O–H distances. In this iteration, the NNP-NEMD simulation was performed by rapidly compressing the system to 70% of its original volume over 10 ps with a 0.5 fs time step. Five independent trajectories were generated at temperatures ranging from 300 to 700 K in 100 K increments, with the structures extracted every 5 fs. After that, 500 structures to be labeled were selected by combining the two screening methods. The numbers of sampled, screened, and labeled structures for iteration 11 are detailed in the last row of **Table 1**. Although the structural feature-based screening method used for the first 10 iterations effectively selects diverse structures, it does not explicitly consider the O–H distance. To address this limitation, we applied a 3D structural feature-based screening method that expands the 2D feature space to a 3D space by incorporating normalized minimum O–H distance values. To evaluate the effectiveness of this approach, 2D and 3D structural feature-based screening methods were employed, each selecting 500 structures.

**Figure 7** shows the results of the 2D and 3D structural feature-based screening methods visualized as 3D spaces. From the lower right figures in **Figure 7(a)** and **(b)**, it is evident that the 3D structural feature-based screening method effectively selects more structures with shorter O–H distances. **Figure 8** further illustrates the impact of this approach by showing the intramolecular O–H distances in the data sets. The data generated by the NNP-MD simulations in the NVT and NPT ensembles (up to iteration 10) contained few structures with short O–H distances despite the large quantity of data. The NNP-NEMD simulations enabled us to sample structures with short O–H distances despite a small quantity of data. Moreover, the 3D structural feature-based screening method further enhanced this effect, particularly in the range of 0.87 to 0.9 Å, where the counts differ by a factor of two to four compared to the 2D structural feature-based screening method.

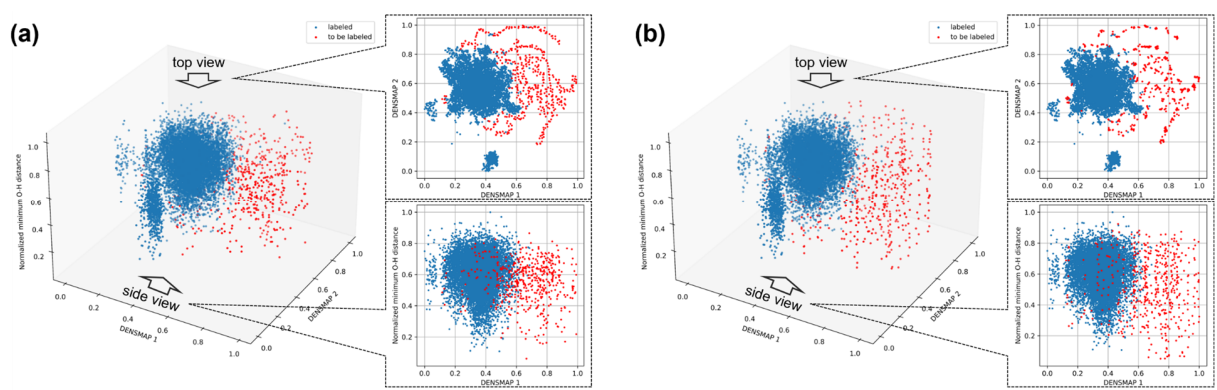

**Figure 7. (a)** 2D and **(b)** 3D structural screening results. They are visualized as a 3D space (left) and as 2D spaces from the top (upper right) and the side (lower right).

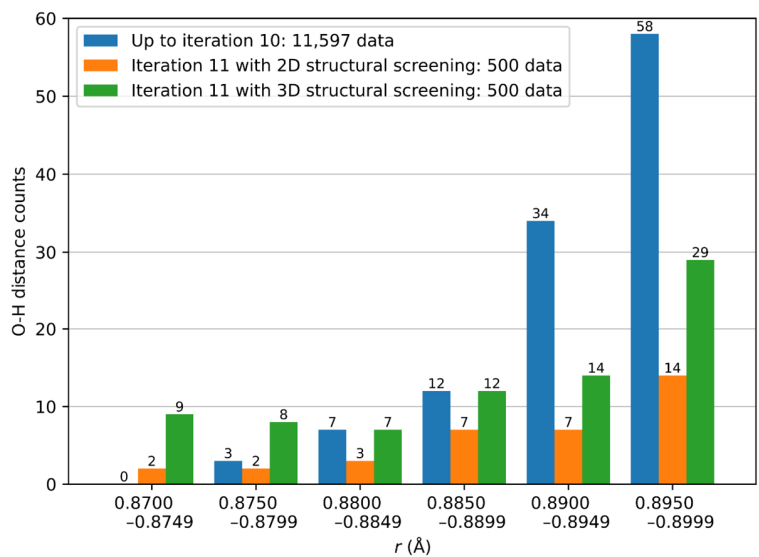

**Figure 8.** Distribution of intramolecular O–H distances in the data up to iteration 10 (blue bar), and iteration 11 with 2D (orange bar) and 3D structural feature-based screening method (green bar). Due to the large number of structures with values above 0.9 Å, this histogram only includes values up to 0.9 Å.

To evaluate the effectiveness of this approach, we ran the simulation again for 1 ns for equilibration and 20 ns for the production run under the same conditions as previously described. **Figure 5(b)** and **(c)** show the densities and maximum model deviations of the NNP-MD simulations using the four models at iteration 11 with the 2D and 3D structural feature-based screening methods, respectively. Although, as shown in **Figure 5(a)** the NNP-MD simulations using the models up to iteration 10 exhibited early collapse in two cases (models 0 and 2). Those with iteration 11 using the 2D structural feature-based screening method (**Figure 5(b)**) appeared to be more stable. Nevertheless, the maximum model deviations in the two cases still increased, leading to collapse at 13 and 20 ns (models 2 and 3). In contrast, all the simulations using the 3D structural feature-based screening method (**Figure 5(c)**) were completed successfully with lower maximum model deviations. This observed stability trend is directly correlated with the number of short O–H distance structures in the data set. This correlation suggests that the repulsive forces between these atoms prevent the system from reaching physically impossible short O–H distances, which would otherwise lead to simulation collapse. Hence, the NNP-NEMD simulations and the 3D structural feature-based screening method were effective in improving the stability of the NNP-MD simulations. We note that the performance of the model 3 in **Figure 5(b)** is much lower than that in **Figure 5(a)**. This is due to the order of training data loading, which affects the training results even when the same initial weights are used. This can result in the NNPs obtained at iteration 11 with 2D feature-based screening is not robust. In contrast, the 3D structural feature-based screening adds data with many short interatomic distances, likely ensuring the robustness of NNPs regardless of the initial weights or data loading order. In addition to the results presented here, further validation of energy conservation in NNP-MD simulations in the NVE ensemble is provided in **Figure S6(a)** in the Supporting Information.

### 3.1.4. Evaluation of the Physicochemical Properties

To measure the accuracy of the generated NNP models, we evaluated the thermodynamic and dynamic properties such as density ($\rho$), self-diffusion coefficient ($D$), thermal expansion coefficient ($\alpha_P$), and isothermal compressibility ($\beta_T$). The thermal expression coefficient and isothermal compressibility can be calculated using the following equations[69]:

$$\alpha_P = \frac{\langle \delta V \delta(H + PV) \rangle_{NPT}}{k_B T^2 V}, \tag{3}$$

$$\beta_T = \frac{\langle \delta V^2 \rangle_{NPT}}{V k_B T} \tag{4}$$

where $V, H, P, k_B,$ and $T$ are the volume, enthalpy, pressure, Boltzmann constant, and temperature, respectively. The self-diffusion coefficient was estimated from the following equation, based on the Einstein relation[70]:

$$D = \frac{1}{6}\lim_{t\to\infty}\frac{d}{dt}\langle|\mathbf{r}_i(t) - \mathbf{r}_i(0)|^2\rangle \tag{5}$$

where $\mathbf{r}_i$ is the atomic coordinates of atom $i$. $D$ was obtained from the slope of the linear portion of the mean square displacement of all atoms. To assess the statistical uncertainty of the calculated properties, we employed a block averaging method. Blocks of 1 ns were used for the density, thermal expansion coefficient, and isothermal compressibility, while 5-ns blocks were used for the self-diffusion coefficient. The respective property was then calculated within each block. Subsequently, the average and standard deviation of each block value were calculated to estimate the overall uncertainty.

**Table 2.** Densities, self-diffusion coefficients, thermal expansion coefficients, and isothermal compressibility at 293.15 K and 1 bar for liquid PG.

| Method | $\rho$ (g·cm$^{-3}$) | $D$ (10$^{-7}$cm$^2$·s$^{-1}$) | $\alpha_P$ (10$^4$K$^{-1}$) | $\beta_T$ (10$^{-5}$bar$^{-1}$) |
|---|---|---|---|---|
| **Experiment** | 1.036[53] | 2.6[71] | 7.0[72] | 4.7[72] |
| **OPLS**[73] | 1.053 | | | 3.7 ± 0.7 |
| **CHARMM27**[73] | 0.978 | 53.0 ± 2 | | 5.3 ± 0.5 |
| **GAFF**[73] | 1.066 | 0.95 ± 0.05 | | 4.8 ± 0.8 |
| **PG_FFM**[73] | 1.057 | 2.5 ± 0.1 | | 3.8 ± 0.7 |
| **This study** | 1.061 ± 0.018 | 5.1 ± 0.4 | 6.6 ± 1.2 | 4.2 ± 0.4 |

Table 2 presents a comparison of the densities, self-diffusion coefficients, thermal expansion coefficients, and isothermal compressibility of the liquid PG obtained from experimental measurement[53,71,72], various MD simulations using traditional force fields[73], and our NNP-MD simulation. The density obtained from our NNP-MD simulation is in good agreement with the experimental value, showing a small error of 2.5% and comparable to that of PG_FFM[73], which is manually tuned specifically for PG. Our NNP-MD simulation also accurately predicted the self-diffusion coefficient, while the CHARMM27 force field significantly overestimated it. The values of the thermal expansion coefficient and isothermal compressibility obtained using our NNP are in excellent agreement with the experimental values with errors of 5.7% and 8.5%, respectively. Overall, our NNP provides a good description of the liquid properties of PG. Notably, our NNP, automatically generated by our NNP generator, achieves comparable or even better accuracy with respect to the commonly used classical force fields (OPSL, CHARMM27, and GAFF), as well as to the manually generated classical force field (PG_FFM[73]).

### 3.2. Polyethylene Glycol (PEG)

Here, we examine the capability of our NNP generator for simulating PEG. An NNP was trained on a data set created from a 4-mer system. For the accuracy evaluation, the NNP-MD simulations of the production run were performed on the 4-, 5-, 6-, 7-, 8-, and 150-mer system.

### 3.2.1. Settings and Results of the NNP Generator

We prepared an initial structure of 5 tetra-ethylene glycol molecules (155 atoms) using the RadonPy Python library[74]. We used the same method as described in ref[59]: the ETKDG version 2[75–77] method implemented in the Python library RDKit[78] for structure exploration, the modified GAFF2[79] for geometry optimization, the Butina clustering[80] based on the torsion fingerprint deviation[81] for clustering, and the DFT with the ωB97M-D3BJ functional[82,83] combined with the 6-31 G(d,p) basis set[84,85] for selecting the most stable structure. Subsequently, the atomic charges of the generated structure were calculated using the restrained electrostatic potential (RESP) charge model[86] with a single-point calculation of the HF/6-31G(d) level of theory[87]. Finally, a 21-step compression/decompression equilibration protocol[88] and an MD simulation in the NPT ensemble for 5 ns at 300 K and 1 bar were performed.

All DFT calculations were carried out using the Quantum ESPRESSO package[34–36]. The Becke–Lee–Yang–Parr (BLYP) exchange-correlation functional[64,65] was used. The Hartwigsen–Goedecker–Hutter (HGH) pseudopotential[89] was employed with a plane-wave cutoff of 100 Ry. To account for the dispersion, an empirical D3 dispersion correction[90] was applied. An initial data set was generated using AIMD simulations at temperatures of 300 K and 600 K, each for 1 ps with a time step of 0.5 fs. By extracting snapshots every 1 fs, a total of 2000 structures were collected. We ran 11 iterations with the same settings as those for the PG simulations, except that a time step of 0.5 fs was used in all NNP-MD simulations. The acceptable range of maximum model deviation of atomic forces, $\epsilon(\boldsymbol{x})$, was set as 0.05 to 0.20 eV/Å. Data points added in each AL cycle were split into training and validation sets in a ratio of 8:2. In iteration 11, the target property of the 3D structural feature-based screening method was set to the minimum H–H distances because the NNP-MD simulations using the models trained on the data set up to iteration 10 become unstable when the distance between hydrogen atoms becomes extremely short.

After iteration 11, a total of 10,105 data points were generated (training: 8387 and validation: 1968). The construction of the data set and comparisons of RMSE values across all iterations are detailed in **Table S1** and **Figure S3** in the Supporting Information, respectively. The validation RMSE of the initial model trained on the AIMD training data set was 0.65 meV/atom for energy and 74.2 meV/Å for force. In contrast, the final model achieved RMSE values of 0.99 meV/atom and 55.2 meV/Å for energy and force, respectively. This trend is like the PG case: although the force error decreased, the energy error increased due to the diversity of energy values in the data set (see **Figure S4** in the Supporting Information). **Figure S5** in the Supporting Information shows the screening results for all iterations. This figure demonstrates, as in the PG case, that our NNP generator can select data points from regions not represented in the existing data set.

**3.2.2. Running the Long-Time NNP-MD Simulations**

To observe the stability of the NNPs, we conducted NNP-MD simulations of the production run for 4-mer PEG. The simulation system was created using the RadonPy library[74] with the same method as that for the initial structure, generating a 4-mer system that contains a hundred molecules (3100 atoms). For a more accurate evaluation, we trained four different models with different initialization seeds and ran four NNP-MD simulations using each trained model in the NPT ensemble at 298.15 K and 1 bar.

**Figure 9** presents the densities and maximum model deviations obtained from the NNP-MD simulations using the models trained on iteration 10 (**Figure 9(a)**) and iteration 11 with the 2D and 3D structural feature-based screening methods (**Figure 9(b)** and **(c)**). The results indicate that model 2, trained on the data set at iteration 10, exhibits significant maximum model deviation, reaching 60 eV/Å at approximately 17 ns, leading to the system collapse. Conversely, in the NNP-MD simulation using model 3 trained on the data set at iteration 11 with the 2D structural feature-based screening method, the maximum model deviation increased, and the density slightly decreased. This was due to the hydrogen atom detaching from the molecule at 3.7 ns. In contrast, the NNP-MD simulations using the models trained at iteration 11 with the 3D structural feature-based screening method showed improvement in simulation stability, enabling all models to complete 21-ns simulations without increasing maximum model deviation. These results suggest that our proposed sampling and screening methods are also effective in improving the stability of NNPs for polymer systems. Beyond the results discussed here, additional validation of energy conservation during NNP-MD simulations in the NVE ensemble can be found in **Figure S6(b)** in the Supporting Information.

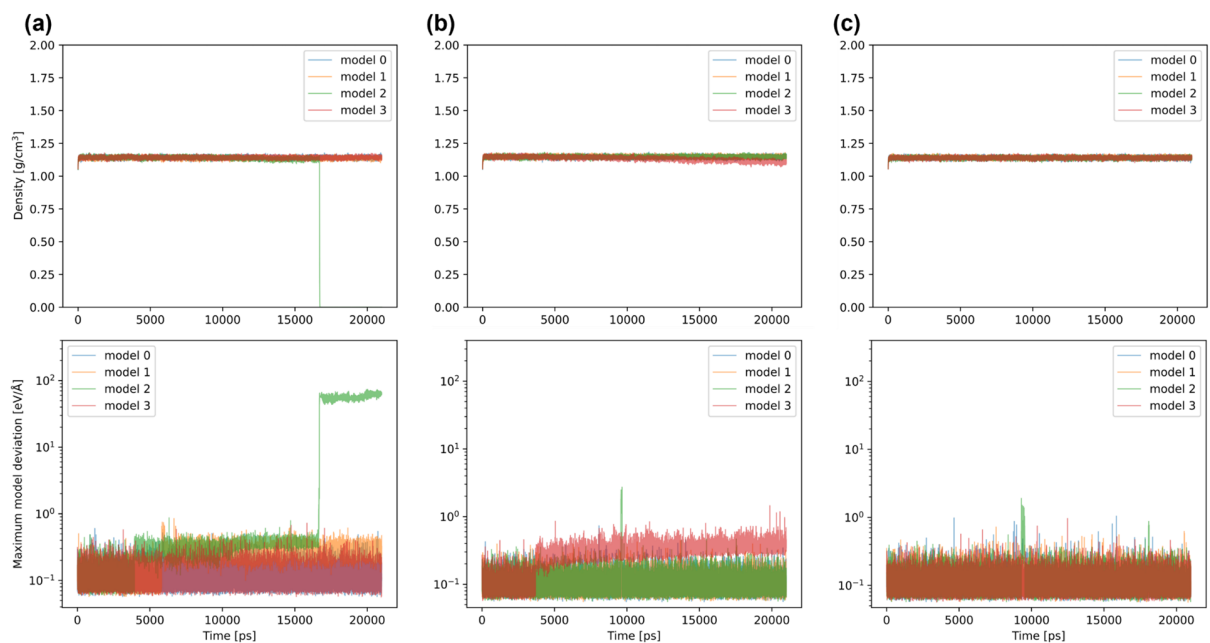

**Figure 9.** Densities and maximum model deviations of atomic forces of simulations PEG at 298.15 K and 1 bar using four models generated at **(a)** iteration 10, **(b)** iteration 11 with 2D structural feature-based screening, and **(c)** iteration 11 with the 3D structural feature-based screening.

### 3.2.3. Evaluation of the Physicochemical Properties

Next, we investigated the accuracy of the density and self-diffusion coefficient predictions for PEG with varying degrees of polymerization. The RadonPy library[74] was also used to create 5-, 6-, 7-, 8-, and 150-mer systems, which contain 3100, 3040, 3150, 3120, 2950, and 10,530 atoms, respectively. The NNP-MD simulations in the NPT ensemble were performed for 1 ns for equilibration and 20 ns for the production run with a time step of 0.5 fs. The temperature was set to 298.15 K and the pressure was set to 1 bar. Using the trained model on the data set at iteration 11 with the 3D structural feature-based screening, all NNP-MD simulations ran stably. From these trajectories, we calculated the density and self-diffusion coefficient.

**Figure 10** presents the equilibrium densities and self-diffusion coefficients obtained from the experimental data[91] at 298.15 K, as well as from OPLS4[92,93], QRNN[93], and this study. The OPLS4 and QRNN results were measured at 300 K. More detailed numerical data are provided in **Table S2** and **Table S3** in the Supporting Information. Regarding the density, OPLS4 and QRNN show good agreement with the experimental data. Our NNP model also achieves good consistency within 1.8% for the 4-mer system. However, it slightly overestimates the density for the degrees of polymerization not included in the training data set. Nonetheless, the error of the 8-mer system is within 3.0%. For the 150-mer system, the predicted density is 1.17 g/cm$^3$, underestimating the experimental value[94] of 1.20 g/cm$^3$ by 2.6%. Overall, our NNP model reproduces the experimental trend, indicating its robustness in predicting densities across different polymer chain lengths. Furthermore, for the self-diffusion coefficients, our NNP model shows excellent agreement with the experimental values, while the OPLS4 and QRNN models underestimate and overestimate them, respectively. For the 150-mer system, our NNP model predicts a self-diffusion coefficient of $0.884 \times 10^{-7}$ cm$^2$/s.

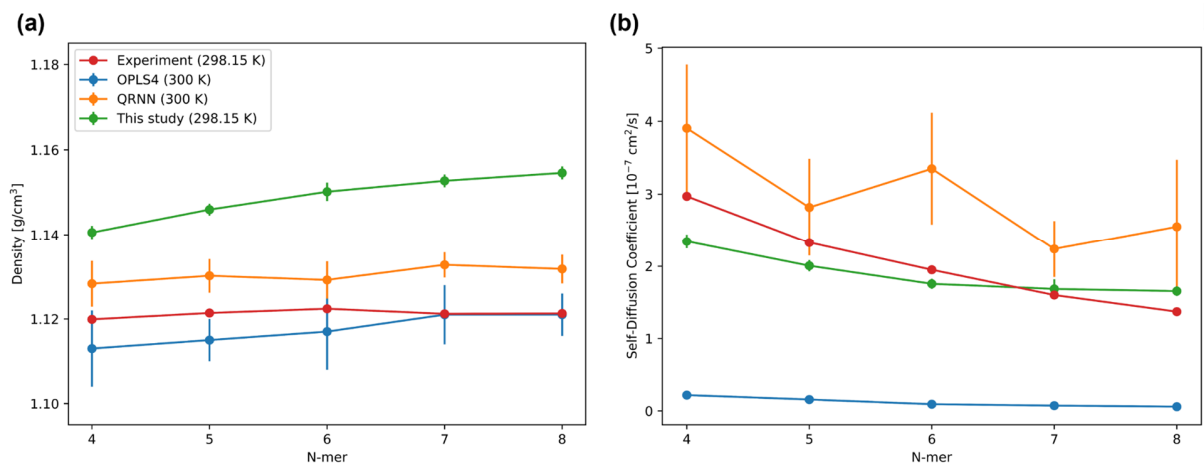

**Figure 10. (a)** Predicted density and **(b)** self-diffusion coefficient from OPLS4[92,93], QRNN[93], and our model compared with the experimental values for PEG.

The close agreements between our NNP model's predictions and the experimental trends for density and self-diffusion coefficients demonstrate its potential for accurately simulating polymer systems. Although slight discrepancies were observed, particularly for longer polymer chains, these findings suggest avenues for further refinement. Expanding the training data set to include a wider range of chain lengths could enhance the model's predictive accuracy. This work underscores the importance of comprehensive data sets in the development of robust and accurate NNPs for simulating complex materials.

The current automatic NNP generator lacks an integrated automatic convergence criterion, which limits the full utilization of AL advantages. In addition, we have not reached a definitive conclusion on the optimal ratio of stable to unstable structures required for producing robust NNPs. As highlighted in recent studies[11,29], ensuring the accuracy of physical property predictions and the robustness of NNP models often requires extensive NNP-MD simulations. These simulations, while essential, can be prohibitively expensive. Addressing this issue is crucial for enhancing the efficiency of our NNP generator. We are committed to refining our methodologies to improve this aspect of our work. Additionally, the current 3D structural feature-based screening present challenges in selecting which element pairs to analyze, thereby affecting the ease of use of our NNP generator. The development of an automatic and efficient method for selecting element pairs could significantly streamline the process and enhance the usability of our system. This advancement would not only improve the user experience but also facilitate more accurate and efficient model generation, ultimately contributing to the broader applicability and effectiveness of our NNP generator.

## 4. Conclusions

We proposed an NNP generator to obtain diverse data for training a robust and accurate NNP. The multiphase workflow includes the initial data set creation, the structure generation by NNP-MD simulations, the combination of the model ensemble-based and structural feature-based screening methods, and the training and evaluation of the NNP models. To address the stability issue of NNP-MD simulations, we developed a solution that combines NNP-NEMD simulations to capture unstable structures with short atomic distances and a 3D structural feature-based screening method that prioritizes these distances while also considering structural feature differences.

The effectiveness of our approach was assessed for liquid PG and PEG. The generated NNPs, trained on the data set generated by the NNP-MD simulations in the NVT and NPT ensembles, initially resulted in the collapse of production runs. The cause of the instability was identified as insufficient data points with short O–H distances for PG and short H–H distances for PEG. To solve this issue, we generated these data using the NNP-NEMD simulations and the 3D structural feature-based screening method. The enhanced NNPs trained on this additional data exhibited significant improvements in the stability of the NNP-MD simulations for both systems. Furthermore, for PG and PEG, we examined the accuracy of the generated NNPs for the thermodynamic and dynamic properties such as density, self-diffusion coefficient, thermal expansion coefficient, and isothermal compressibility. These NNPs accurately predicted these properties as observed in experiments, achieving comparable or even better accuracy than the commonly used and manually generated classical force fields.

These findings establish a methodology for developing robust and accurate NNPs for organic materials using our NNP generator. These robust and accurate NNPs enable the execution of nanosecond-scale stable NNP-MD simulations of large systems containing >10,000 atoms. Furthermore, our automated NNP generator greatly reduces the time required for NNP generation, allowing researchers to allocate more time to further scientific investigations and potentially accelerating the discovery and development of new materials. Although this study focused on organic systems, the same workflow can be applied to inorganic systems, suggesting the broad applicability of our approach across different material types.

# Supporting Information

## 1. Analysis of the NNP Generation Process for Propylene Glycol (PG)

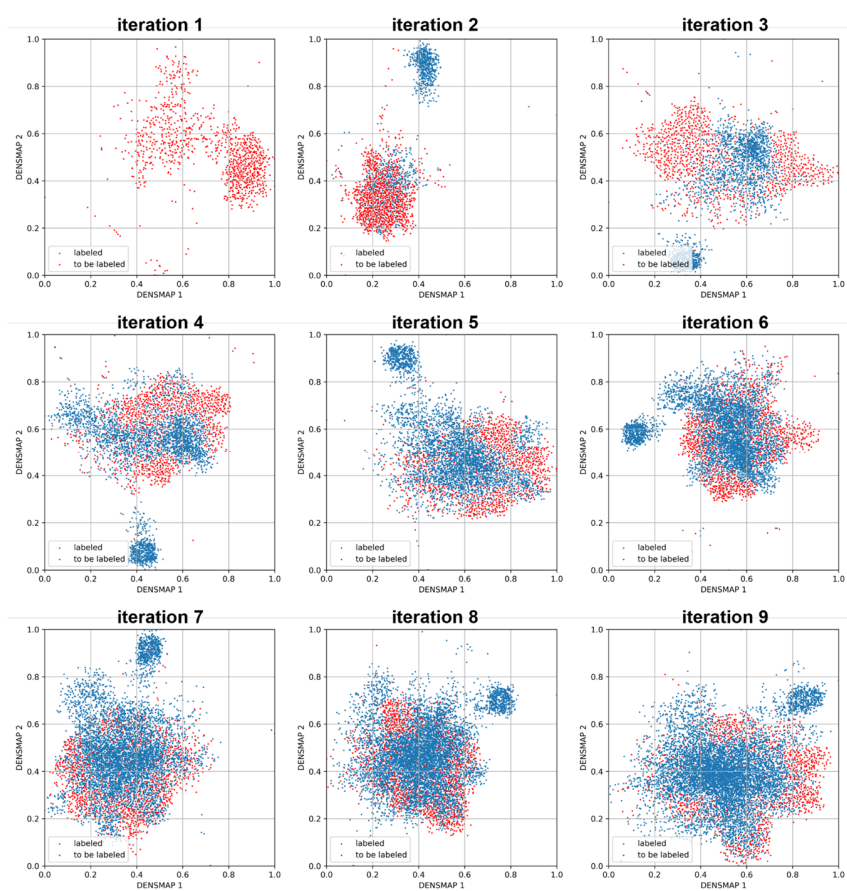

**Figure S1.** Two-dimensional feature spaces after screening for PG at each iteration. The isolated dense group seen in iteration 2 originated from iteration 1, generated by NNP-MD simulations using an NNP model trained on only a 1 ps AIMD simulation. These data points show a distinct trend from those in later iterations. As shown in the upper right figure of

**Figure 7(b)**, structures from the NNP-NEMD simulation, which have shorter interatomic distances, occupy a similar isolated feature space as the dense group from iteration 1. This indicates that the dense group contains more unstable structures than those from iterations 2 to 10.

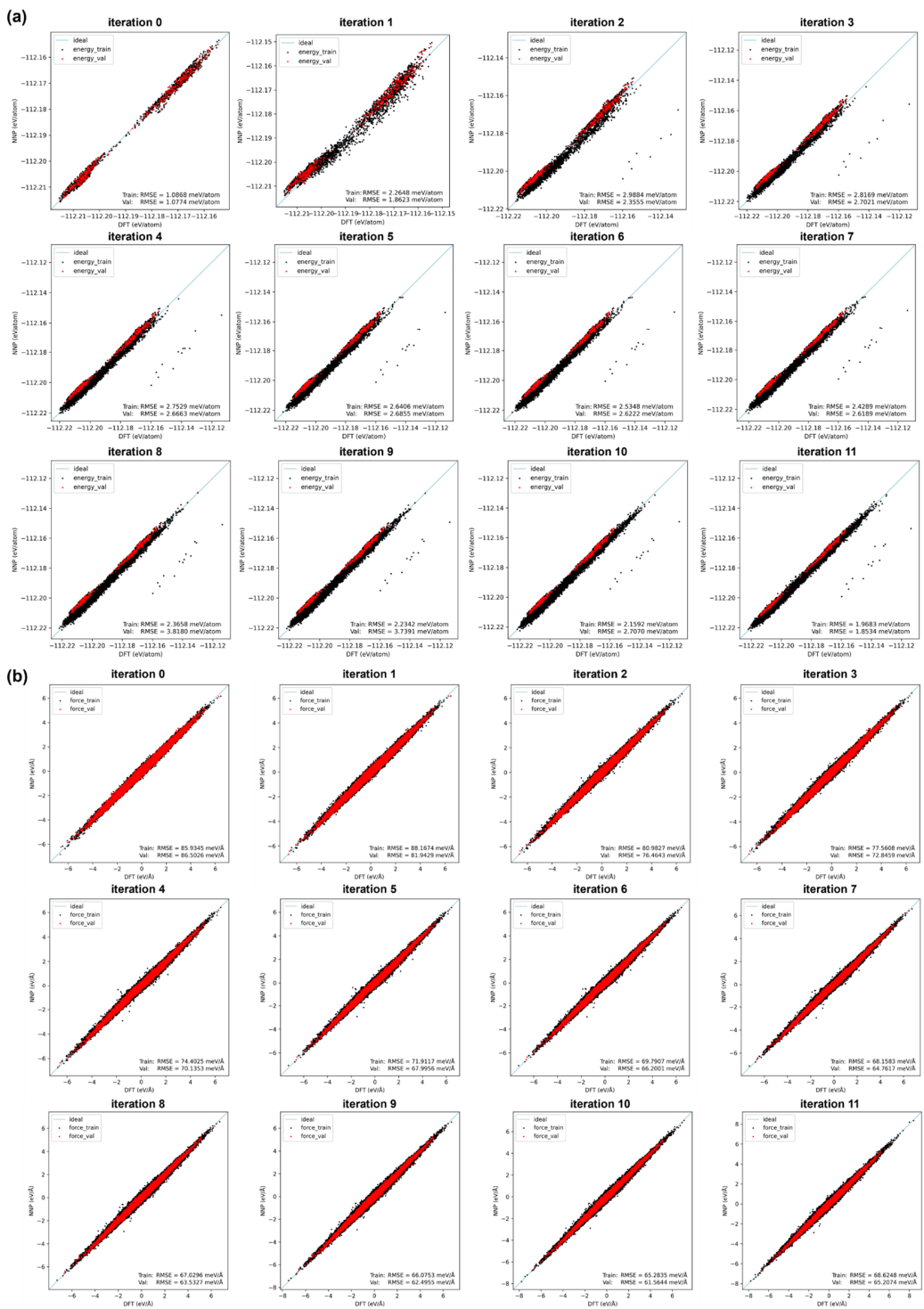

**Figure S2.** Comparison of DFT and NNP at each iteration for PG: **(a)** energy and **(b)** force predictions.

## 2. Analysis of the Neural Network Potential Generation Process for polyethylene glycol (PEG)

**Table S1.** Construction of the data set for PEG[a]

| iter | sampling | | | | screening | | labeling |
|---|---|---|---|---|---|---|---|
| | method | ensemble | temperature (K) | # of sampled structures | # of structures selected by the model ensemble-based method | # of structures selected by the structural feature-based method | # of labeled structures |
| 0 | AIMD | NVT | 300, 600 | 2000 | | | 2000 |
| 1 | NNP-MD | NVT, NPT | 300–600 | 12,167 | 903 | 903 | 903 |
| 2 | NNP-MD | NVT, NPT | 300–600 | 14,000 | 12,517 | 1000 | 1000 |
| 3 | NNP-MD | NVT, NPT | 300–600 | 13,648 | 12,996 | 1000 | 999 |
| 4 | NNP-MD | NVT, NPT | 300–600 | 14,000 | 12,568 | 1000 | 998 |
| 5 | NNP-MD | NVT, NPT | 300–600 | 14,000 | 13,291 | 1000 | 998 |
| 6 | NNP-MD | NVT, NPT | 300–600 | 9817 | 9120 | 1000 | 957 |
| 7 | NNP-MD | NVT, NPT | 300–600 | 14,000 | 12,963 | 1000 | 1000 |
| 8 | NNP-MD | NVT, NPT | 300–600 | 14,000 | 13,085 | 1000 | 1000 |
| 9 | NNP-MD | NVT, NPT | 300–600 | 14,000 | 11,495 | 1000 | 1000 |
| 10 | NNP-MD | NVT, NPT | 300–600 | 13,042 | 10,811 | 1000 | 1000 |
| 11 | NNP-NEMD (compress to 70%) | | 300–600 | 14,000 | 11,537 | 500 | 500 |

[a]Each row details the additions to the training data set at each iteration. The columns represent iteration, sampling method, ensemble, temperature, number of sampled structures, number of screened structures by the model ensemble-based method, number of screened structures by the structural feature-based method, and number of labeled structures. The pressure of the NNP-MD simulations in the NPT ensemble was set to 1 bar.

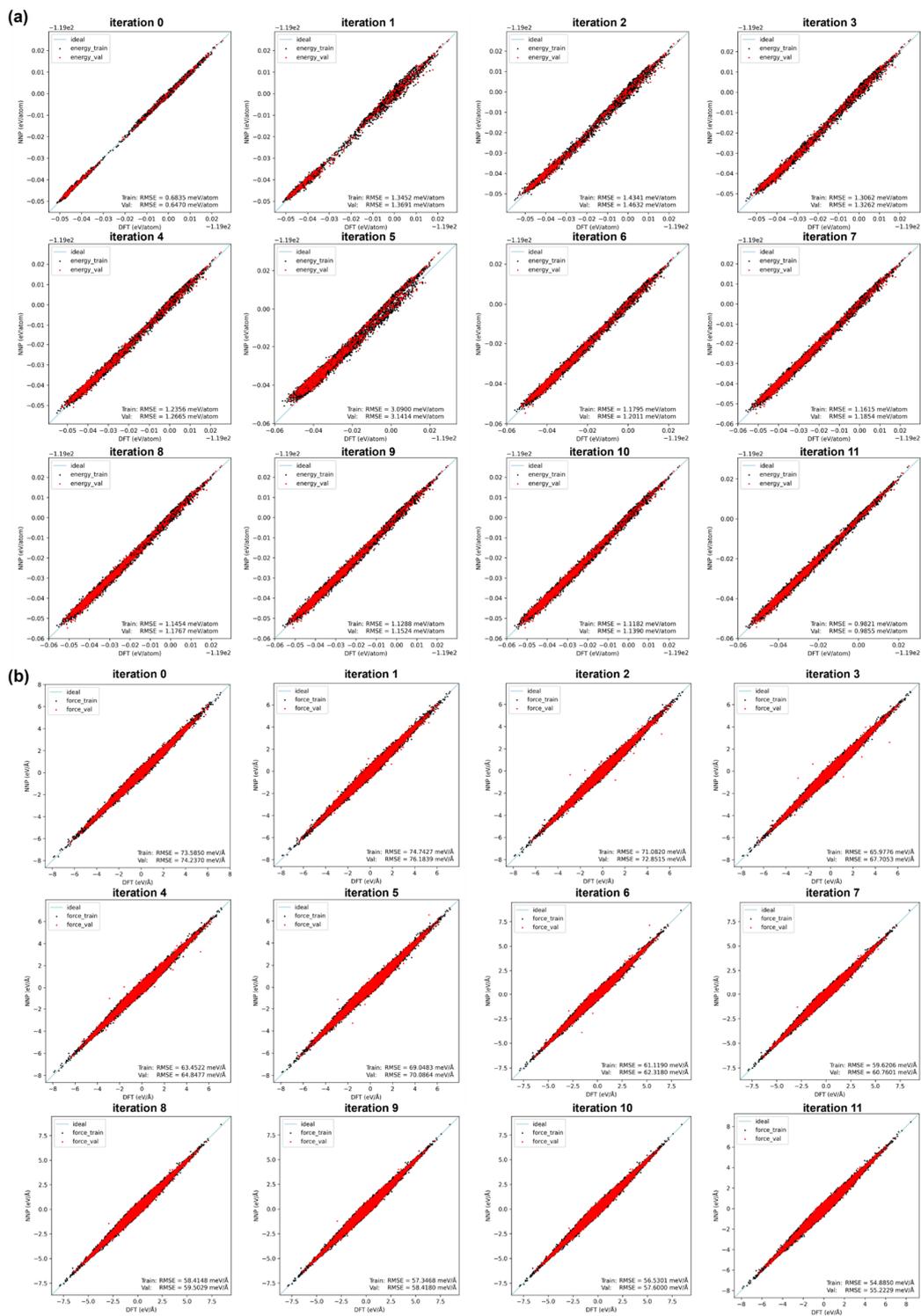

**Figure S3.** Comparison of DFT and NNP at each iteration for PEG: **(a)** energy and **(b)** force predictions.

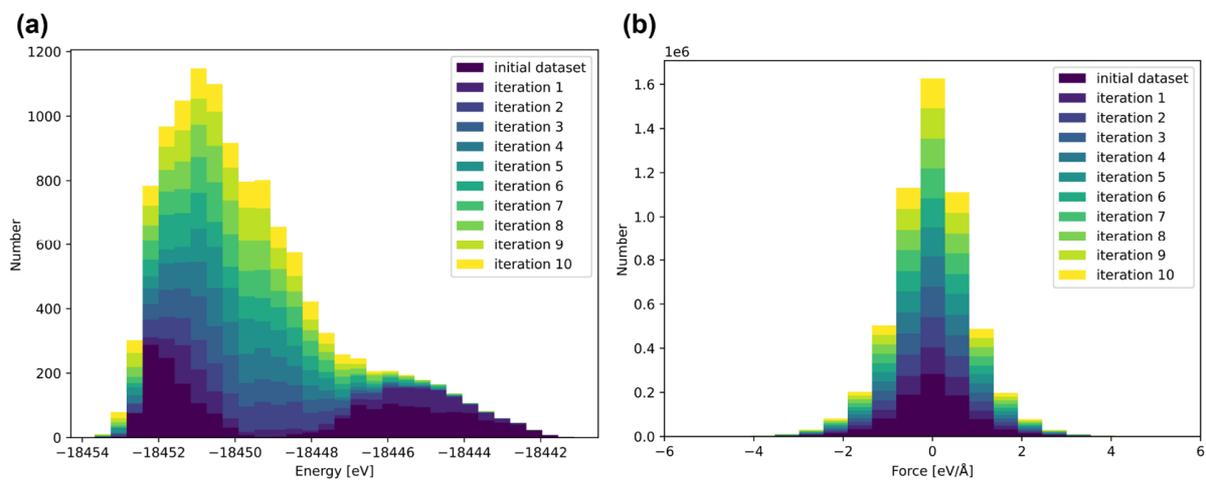

**Figure S4.** Histograms for the data sets of PEG over 10 iterations for (a) energy and (b) force. The energy histogram of the initial data set (dark blue bars) exhibits two distinct peaks corresponding to 300 K (from −18,543 to −18,450 eV) and 600 K (from −18,448 to −18,442 eV).

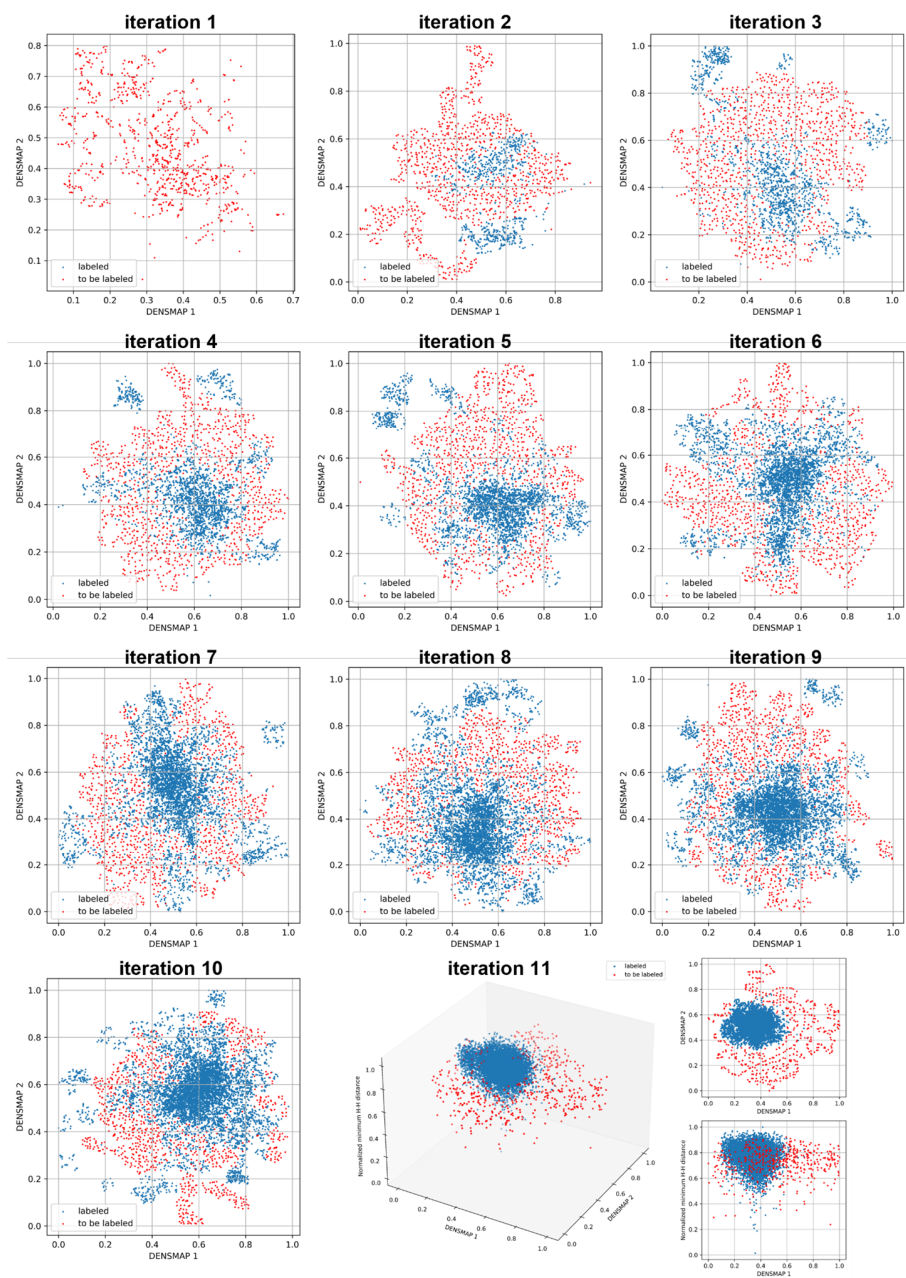

**Figure S5.** Feature spaces after screening: two-dimensional (iterations 1–10), and three-dimensional (iteration 11) for PEG. The broader sampling space compared to PG is due to system complexity. **Figure S1** shows feature maps of PG with 13 atoms per molecule, limiting structural diversity and constraining sampling space. In contrast, this figure depicts feature maps of PEG with 31 atoms per molecule, enabling greater structural diversity and a wider sampling space.

## 3. Comparison of the simulation results with the experiments for polyethylene glycol (PEG)

**Table S2**. Densities of PEG at 298.15 K and 1 bar: experimental vs. simulation.

| Polymer chain length | Density (g/cm³) | | | |
|---|---|---|---|---|
| | OPLS[1,2] (300 K) | QRNN[2] (300 K) | This study (298.15 K) | Experiments[3] (298.15 K) |
| 4 | 1.113 ± 0.009 | 1.128 ± 0.005 | 1.141 ± 0.001 | 1.120 |
| 5 | 1.115 ± 0.005 | 1.130 ± 0.004 | 1.146 ± 0.001 | 1.121 |
| 6 | 1.117 ± 0.009 | 1.129 ± 0.004 | 1.150 ± 0.001 | 1.122 |
| 7 | 1.121 ± 0.007 | 1.133 ± 0.003 | 1.153 ± 0.001 | 1.121 |
| 8 | 1.121 ± 0.005 | 1.132 ± 0.003 | 1.155 ± 0.001 | 1.121 |

**Table S3**. Self-diffusion coefficients of PEG at 298.15 K and 1 bar: experimental vs. simulation.

| Polymer chain length | Self-diffusion coefficient ($10^{-7}$ cm²/s) | | | |
|---|---|---|---|---|
| | OPLS[1,2] (300 K) | QRNN[2] (300 K) | This study (298.15 K) | Experiments[3] (298.15 K) |
| 4 | 21.8 ± 3.1 | 390.4 ± 87.3 | 234.6 ± 9.6 | 297 |
| 5 | 15.7 ± 5.8 | 281.7 ± 66.7 | 200.6 ± 7.7 | 233 |
| 6 | 9.2 ± 2.5 | 334.8 ± 76.8 | 175.5 ± 7.0 | 195 |
| 7 | 7.3 ± 2.3 | 223.9 ± 38.9 | 168.4 ± 13.3 | 160 |
| 8 | 5.8 ± 1.3 | 255.2 ± 92.0 | 165.4 ± 6.7 | 137 |

## 4. Validation of Energy Conservation in NNP-MD Simulations

To verify whether our NNP models adhere to the energy conservation law, we conducted NNP-MD simulations in the NVE ensemble for 1 ns with a timestep of 0.5 fs. The velocities were initialized at a temperature of 298.15 K. **Figure S6(a)** and **(b)** show the total energy, potential energy and kinetic energy for PG and PEG, respectively. It is evident that the total energy remains constant, while the potential energy and kinetic energy exhibit fluctuations. Thus, we can conclude that the energy conservation law is satisfied by our NNP models.

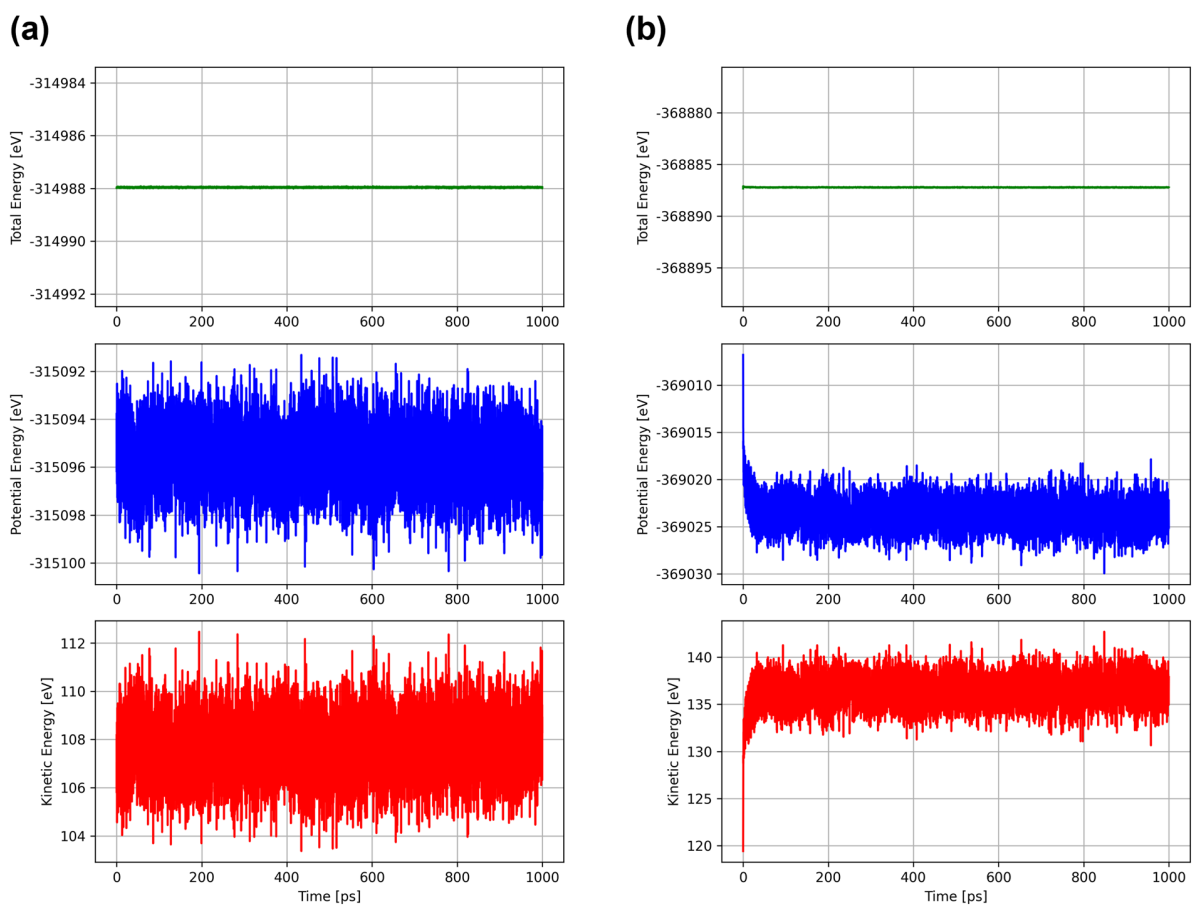

**Figure S6.** Total energy (top), potential energy (middle) and kinetic energy (bottom) for (a) PG and (b) PEG from NNP-MD simulations in the NVE ensemble.